\documentclass[prl,aps,twocolumn,preprintnumbers,showpacs,superscriptaddress]{revtex4}
\usepackage[colorlinks=true, pdfstartview=FitV, linkcolor=red, citecolor=blue, urlcolor=blue]{hyperref}

\usepackage{graphics,graphicx,epsfig,bm}
\usepackage{epstopdf}

\newcommand{\ket}[1]{| #1 \rangle}

\newcommand{\nv}{N$-V$}
\newcommand{\nfourteen}{$^{14}$N}
\newcommand{\nfifteen}{$^{15}$N}
\newcommand{\cthirteen}{$^{13}$C}

\newcommand {\Fig}[1] {Figure~\ref{#1}}

\begin{document}

\title{Brokered Graph State Quantum Computing}

\date{\today}

\author{Simon C. Benjamin}
\affiliation{Department of Materials, Oxford University, Oxford OX1
3PH, United Kingdom}
\affiliation{Centre for Quantum Computation, Department of Physics,
Oxford University OX1 3PU, United Kingdom}

\author{Dan E. Browne}
\affiliation{Department of Materials, Oxford University, Oxford OX1
3PH, United Kingdom}
\affiliation{Centre for Quantum Computation, Department of Physics,
Oxford University OX1 3PU, United Kingdom}

\author{Joe Fitzsimons}
\affiliation{Department of Materials, Oxford University, Oxford OX1
3PH, United Kingdom}
\affiliation{Department of Mathematical Physics, NUI Maynooth, Maynooth, Ireland}

\author{John J. L. Morton}
\affiliation{Department of Materials, Oxford University, Oxford OX1
3PH, United Kingdom}
\affiliation{Centre for Quantum Computation, Department of Physics,
Oxford University OX1 3PU, United Kingdom}


\begin{abstract}
We describe a procedure for graph state quantum computing that is tailored to fully exploit the physics of optically active multi-level systems.
Leveraging ideas from the literature on distributed computation together with the recent work on probabilistic cluster state synthesis, our model assigns to each physical system two logical qubits: the {\em broker} and the {\em client}. Groups of brokers negotiate new graph state fragments via a probabilistic optical protocol. Completed fragments are mapped from broker to clients via a simple state transition and measurement. The clients, whose role is to store the nascent graph state long term, remain entirely insulated from failures during the brokerage. We describe an implementation in terms of \nv~centres in diamond, where brokers and clients are very naturally embodied as electron and nuclear spins. \end{abstract}

\maketitle

\section{Introduction}
In the field of quantum information, the beautiful concept of graph state computation is blossoming~\cite{blossom}. The term {\em graph state}, a generalization of the more common term {\em cluster state}, refers to a certain kind of multi-qubit entangled state which one would prepare prior to the computation~\cite{RandB, BandR, Long,Graphs}. This state has the property that the computation can then proceed purely by single-qubit measurement, consuming the entanglement as a resource in a so called {\em one-way computation}. Thus one introduces a degree of separation between the act of creating entanglement and the act of executing the computation, which promises to be a tremendous advantage in many implementations; Refs.\ \cite{nielsenLOQC,Jaksch,browneLOQC,Kay} are a few examples. 

Several recent publications \cite{BK,BenjComment,Almut,BES,LimLong} have explored the appealing possibility of graph state generation using matter qubits, with the obvious benefits that they are static and potentially long lived, together with an optical coupling mechanism to create suitable entanglement. The matter qubits can be completely separate, for example each within its own cavity apparatus, providing that a suitable optical channel connects them. These exciting ideas all involve a single qubit in each cavity, which causes them to suffer a generic problem. When the inter-cavity entanglement operation fails -- due either to an inherent  indeterminism or to photon loss, for example -- then the nascent graph state is damaged. The two qubits, which may already be high value nodes in the graph, now have an undefined state. They must be removed and the corresponding sections must be regrown.

\begin{figure}[!t]
\centering
\includegraphics[width=7.cm]{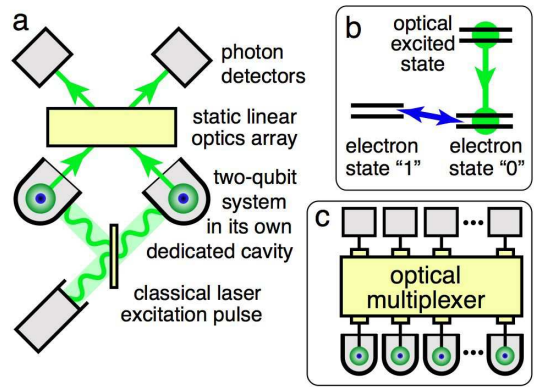}
\caption{ (a) Schematic of the basic apparatus: individual atoms (or atom-like systems such as \nv~centres in diamond) are each isolated within a {\em separate} cavity. Each system has a multi-level eigenspectrum within which one can define two qubits. They are the \emph{broker} (which would typically be represented by electron spin states) and the \emph{client} (requiring long decoherence times and being naturally associated with a nuclear spin). Each cavity apparatus includes mechanisms, such as ESR/NMR pulse generation, that implement deterministic local \emph{broker-client} interactions as level transitions. The broker qubits associated with different atoms can be entangled via the emission of a photons into a `path erasure' optical apparatus. Clients are insulated from broker failures (see Fig.~\ref{fig:scheme}).   (b) Simplified view of one possible level structure (elaborated in Fig.~\ref{elevels}). The electronic state can be mapped into a photon via an optically-allowed transition. Doublets correspond to the two nuclear states. (c) In a mature form of the technology, multipartite brokering entangles arbitrary sets of broker qubits in parallel, via an optical multiplexer. Related devices already exist \cite{BellLambda}.} \label{fig:intro}
\end{figure}

The impact of a low probability $p$ of successfully entangling the matter qubits is that the graph may suffer repeated damage prior to each positive growth event. Remarkably, this can be tolerated in the sense that systematic average growth is always possible with a suitable strategy \cite{BK,BenjComment}, but for low $p$ the growth rate becomes extremely slow. Meanwhile there is a substantial earlier literature on distributed QIP over imperfect channels. This established the utility of having at least two logical qubits within each locality so that one can remain insulated from randomness suffered by the other (see Refs.\ \onlinecite{grover, jens1,jens2, sandu}), a insight that lead to implementation schemes in the context of ion traps, for example~\cite{ionTrapCoupling}. Here, we unite and extend ideas from these publications in order to form a broker-client model for graph state quantum computing. This model is tailored to exploit the established physics of optically active electron-nuclear systems, such as \nv~centres in diamond, in such a way as to provide deterministic graph growth despite non-deterministic entangling operations. 

General graph states, which may contain nodes of high degree (i.e. a high number of attached edges), can then be formed without putting such nodes at risk of damage though entanglement failure. Thus we have no need to restrict ourselves to cluster states, which are graph states with a regular, e.g. square, lattice. A cluster state supporting a given function can generally be reduced to a graph state capable of the same function but requiring far fewer qubits~\cite{Graphs}. There may also be more profound advantages; for example, it has been shown that certain graph state structures give robustness against error, for example qubit loss \cite{losstolerant} and particular layouts are key to general  fault tolerant graph state computation \cite{faulttolerant}.

\begin{figure}[!t]
\centering
\includegraphics[width=8.3cm]{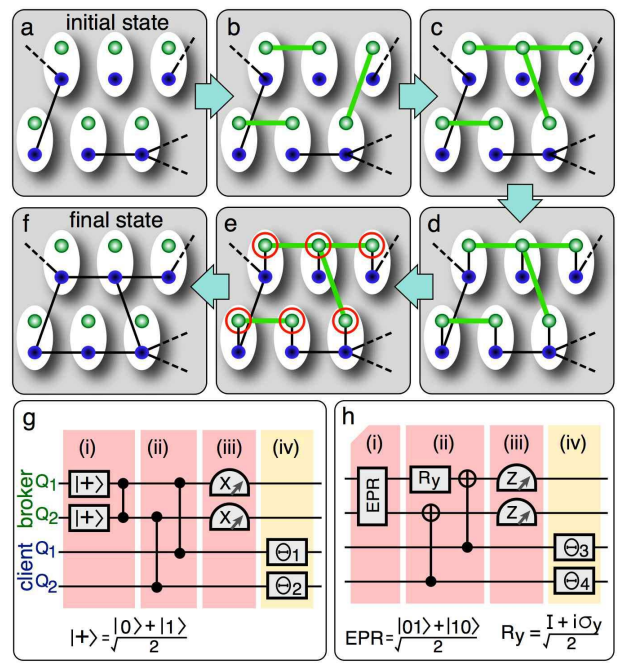}
\caption{ The procedure for brokering graph states. The ovals indicate the client (below, blue) and broker (above, green) qubits associated with a single physical system in a dedicated cavity. (a) We are free to assume that an arbitrary set of graph edges, i.e. control-phase gates, have already been implemented as shown. (b) \& (c) Where we wish to create new graph edges, we first create entanglement in the broker degree of freedom. This may take many attempts, and may involve reseting the electronic states. Here we depict two graph fragments brokered independently: the simple bipartite fragment (lower left) and a more complex `3-node' fragment formed in a two stage process involving four brokers. (d) We then deterministically create graph edges between the brokers and their client qubits {\em within} each pair. (e) Finally, we project the broker entanglement onto their client qubits via measurements on the brokers. (f) The clients thus acquire new entanglement relations with one another.
(g) A naive circuit diagram for the simplest bipartite brokering involves preparation state $\ket{+}$, phase gates and $X$ measurement. (h) An optimized equivalent circuit can be constructed from realistic fast operations including control-NOT and $Z$ basis measurements. The $Z$ rotations on the client qubits, marked $\Theta$,  can be postponed and subsumed into the eventual measurement procedure.
}
\label{fig:scheme}
\end{figure}

\section{Brokered graph state formation}

We suggest that each cavity should contain an atom (or atom-like system such as an \nv~centre) that has a sufficiently complex level structure to constitute two logical qubits: the broker and the client (illustrated in Fig.\ \ref{fig:intro}). We assume that we can perform deterministic quantum gates between these two, in contrast to the inter-broker operations which may be error prone and inherently indeterministic. For many systems, the natural embodiment would be an electronic qubit for the broker, and a nuclear qubit for the client. Nuclear states are ideal for storing entanglement during the primary graph's growth period, due to low decoherence rate; moreover the electron-nuclear hyperfine interaction would permit selective  transitions corresponding to, e.g. controlled-not operations between client and broker. Suitable physical systems have already been well explored experimentally, as we presently discuss. In this way, we effectively place a two-qubit quantum computer within each cavity, which has profound consequences for our strategy for creating entanglement {\em between} cavities. We can now assign the broker qubits to the entanglement generation process, while the nuclear qubits embody the graph state itself (see \Fig{fig:scheme}). The broker states can be reset or projected into a given basis {\em without} harming the nascent graph. Similarly, failures in the optical entanglement process (whether inherent indeterminism or errors such as photon loss) only retard the creation of the brokered graph fragment and {\em do not} damage the primary graph held by the client qubits. This is of course assuming that such errors are {\em heralded}, as we presently discuss.

\section{Multipartite brokerage}

Brokerage is not be restricted to the bipartite level, i.e. creating single graph edges prior to mapping those edges onto the clients. The graph state approach can be exploited by allowing the broker qubits to arrange more complicated multipartite graph fragments, before mapping this onto the client qubits. The question of whether one should aim to create such larger fragments in the broker space depends on the experimental parameters, as we now illustrate using the example of Fig.\ \ref{fig:scheme}(b)\&(c) where a 3-node (a central node connected to three outer vertices) is established.

If the broker qubits had been restricted to generating only EPR pairs, then it is easy to see that three complete rounds of brokering and transfer to clients would be needed to create the desired client entanglement \Fig{fig:scheme}(f). The approximate time required will be

\[
\tau_{sequential} = 3(\frac{1}{p}(\tau_H + \tau_{O} + \tau_M) + \tau_{CNOT} + \tau_H)\,\,,
\]
where $\tau_H$, $\tau_{CNOT}$, $\tau_M$ are the times required to implement a broker Hadamard gate, a CNOT controlled by the client qubit, and a measurement in the broker computational basis. The average time spent on a single attempt to generate optical entangled is $\tau_{O}$. If, however, the broker qubits are themselves used to create a 3-node before being entangled with the client qubit, as depicted in \Fig{fig:scheme}, then the time required will be

\[
\tau_{3-node} = \frac{1}{p}(\frac{1}{p}(\tau_H + \tau_O + \tau_M) + \tau_O) + \tau_{CNOT} + \tau_H \,\,.
\]

For a typical physical implementation in terms of electron brokers and nuclear clients, we would anticipate that $\tau_{O} \approx \tau_H$ (i.e. the electron rotations are the slow part of the optical entanglement) and $\tau_M \ll \tau_H$ (since measurement is purely optical). It is then advantageous to use the latter scheme if the inequality, $\frac{\tau_{CNOT}}{\tau_H} \ge \frac{1}{p^2} - \frac{5}{2p} - 1$,
holds. For experiments in the near future, it is reasonable to take probability $p=0.25$. We may take ratio $\frac{\tau_{CNOT}}{\tau_H} = 10$ at least, due to the need to implement the CNOT with a frequency selective pulse while unconditional electron rotations can correspond to `hard' pulses. With these vales there is already an advantage in brokering the larger fragment: $\tau_{brokered} \approx \frac{2}{3}\tau_{sequential}$, and approaches $\frac{1}{3}\tau_{sequential}$ as this ratio grows. Higher values of $p$, e.g. from experimental refinements reducing photon loss, would increase this advantage. Thus, we conclude that it may indeed be desirable to create mutual entanglement between four brokers. For more extreme values of the experimental parameters, it may even be desirable to further increase the degree of multipartite brokerage.

An extension of this model would consider the utility of recruiting redundant brokers, i.e. broker qubits not presently being used by their clients, as an additional resource for rapidly creating sophisticated broker entanglement. It would appear that one should aim to use all of the brokers all of the time, regardless of whether their local clients require new entanglement during the present step of the graph growth. A full exploration of this idea is beyond the scope of the present paper.

\bigskip

\section{Entangling operations for the broker qubits}

A wide variety of options exist for entangling macroscopically separated qubits, and proposals have been made suitable for many different physical implementations. Many proposals for the generation of entanglement between atoms in separate cavities combine the detection of emitted photons with the erasure of which path information to generate maximally entangled states (Bell states) in the atoms  \cite{Cabrillo}. The starting point for these schemes is to prepare the atoms such that they emit a photon into a cavity mode conditional on their internal state. After leaking from the cavity, the photon(s) pass(es) through a beam splitter (which erases which path information) and arrives at detectors. Certain detection outcomes indicate the creation of a maximally entangled two qubit state (Bell state)  between the atoms. Thus although failure may occur, such failures are always known, or {\em heralded} - this is a crucial feature of all the schemes we consider here.

One versatile implementation~\cite{Bose KPV 99,brownehuelga}  employs  a Raman-type transition on a three-level $\Lambda$ configuration of atomic levels  with one transition is coupled non-resonantly to a cavity mode,  the other driven by a non-resonant laser. When the external driving field is weak and continuous \cite{brownehuelga} this approach is robust to many experimental imperfections, does not require atoms to be prepared in a superposition state, requires no photon number resolving detectors and has a high success probability.  Raman-type transitions can be implemented in a variety of systems. In \nv~centre systems such transitions have been proposed \cite{hemansonprb} and demonstrated experimentally \cite{hemmeroptlett}. 

In a system where the broker qubit levels each possess a transition to an excited state via modes of different polarizations, methods using  two-photon polarization interference can also be employed \cite{polarization}. These have the advantage of being robust against path length fluctations between cavities.

The goal of these schemes is the robust preparation of Bell states, typically without preserving entanglement previously associated with the qubits. This would suffice for bipartite brokering in our model. However, there are also schemes by which multipartite brokering may be achieved. Several suitable schemes\ \cite{Almut, LimLong, entgates} could build up fragments via successive pairwise entanglement of brokers.  An alternative powerful approach is to use more general projections (for example projections onto parity sub-spaces) which  have the  effect of ``fusing together''  small graph states (for example Bell states) so that larger graph states of arbitrary layout can be formed \cite{BES,verstraete,browneLOQC}. A particularly robust method of implementing fusion operators in the cavity / matter qubit setting was  proposed recently\ \cite{BK,BenjComment}. This last scheme, in particular, fully supports the preceding model together with the following \nv~centre implementation. 

\bigskip
\section{A physical system: \nv~centres in diamond}

We have remarked that a natural implementation of the broker-client model could employ a pair of coupled electron and nuclear spin, such as can be found in atoms, or defects in solid state samples. Single electron spins can be detected optically~\cite{jelezko04}, enabling remote entanglement as described above, while nuclear spins can benefit from coherence times as long as 10s of seconds, even at room temperature in the solid state~\cite{ladd05}.  

\begin{figure}[t]
\centering
\includegraphics[width=6 cm]{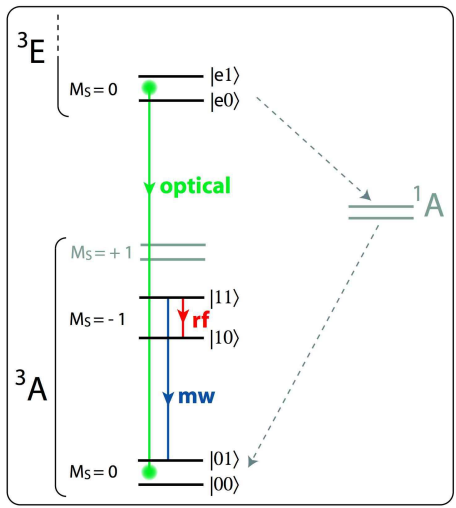}
\caption{ Energy level diagram for a typical \nv~centre in diamond, including electron and nuclear spin states. Relevant qubit states are labeled in the $\ket{broker,~client}$ basis. A small magnetic field is required to split the $M_S=\pm1$ states; this also lifts the degeneracy of the nuclear spin in the $M_S=0$ manifold, where the hyperfine interaction is 0. A selective $\pi$ rotation applied to the transition labeled `mw' corresponds to a CNOT gate targeting the electron spin, controlled by the nucleus (vice versa for the equivalent operation on the transition labeled `rf'). The zero-phonon-line energy of the optical transition is about 1.95~eV ($\lambda \approx$~637~nm).}
\label{elevels}
\end{figure}

While we wish to stress that the requirements of our scheme permit many possible physical implementations, the discussion which follows illustrates a particular embodiment, using \nv~defects in diamond. The defect consists of a substitutional nitrogen atom (\nfourteen~or \nfifteen) and a vacancy in an adjacent site. This contains all the ingredients required by the brokered graph state generation scheme. The ground state of the defect consists of an electron $S=1$ spin triplet ($^3A$) which has a strong dipole allowed optical transition to a first excited spin triplet state ($^3E$), as shown in \Fig{elevels}. The electron spin can be coupled to a range of different nuclei (\cthirteen, \nfourteen~or \nfifteen) via a hyperfine interaction. Depending on which coordination shell the ($I=1/2$) \cthirteen~nucleus inhabits, the strength of the hyperfine interaction can be as high as 200~MHz (nearest neighbour), falling off rapidly for more distant sites~\cite{wrachtrup01, cox94}. Thus, if a \cthirteen~nucleus is chosen as the nuclear (client) qubit, a wide range of choices for the electron-nuclear interaction strength are available. The anisotropy of this interaction also falls with increasing distance from the defect. Alternatively, the coupling to the nitrogen nucleus may be exploited --- this can vary from $2-100$~MHz for \nfourteen~($I=1$), depending on the precise species of \nv~centre~\cite{he92,cox94}, however the spectrum is complicated by the nuclear quadrupolar interaction. The less abundant \nfifteen~avoids this additional splitting, being $I=1/2$, and has a hyperfine constant approximately 40$\%$ greater than that of \nfourteen~\cite{cox94}.

We propose using this optically-detectable electron spin coupled to some $I=1/2$ nuclear spin with an interaction strength ranging between 5 to 200~MHz. The electron spin is initialised in the $M_S=0$ ground state through laser cooling. Entanglement is achieved between the broker (electron) spins by the scheme of Ref.~\cite{BK,BenjComment}, for example. Thereafter, a CNOT operation targeting the electron spin and controlled by the nuclear spin is achieved through a \emph{selective} microwave $\pi$ pulse, as shown in \Fig{elevels}. This establishes entanglement with the client (nuclear) spins. Finally, the electron spin undergoes a rotation, achieved by a hard (non-selective) microwave pulse, and is then optically measured -- projecting the broker's node in its graph fragment onto the client nucleus. Typical microwave pulse lengths are 50~ns, while optical measurement of the electron spin can be faster, given that the free space radiative lifetime of the $^3E$ state of 13~ns~\cite{he92}.

When we wish to consume our graph state to perform the eventual computation, projective measurements of the client qubits are required. In order to achieve this in a nuclear/electron system such as the one we describe here, it will be necessary to map the nuclear state onto the electron for subsequent optical readout. We use a SWAP operation to exchange the client qubit from the nucleus to the electron; it is efficient to take this opportunity to also place a `fresh' superposition on the nucleus, so that it can be reincorporated into the graph as a new client. Specifically, the electron spin is initialised into the $\ket{0}+\ket{1}$ state, then SWAP'ed with the nuclear spin using three CNOT gates. This necessarily requires one CNOT targeting the nuclear spin, which must be performed using the a slower radiofrequency (RF) selective $\pi$ pulse on the nucleus (shown in \Fig{elevels}). A typical RF pulse duration is 10$~\mu$s. The electron spin can then be rotated and measured as described above.

Optical excitation occurs only within the $M_S=0$ subspace, in which the electron-nuclear interaction is zero. Therefore the nuclear spin should remain unperturbed by the optical excitation to an excellent approximation.  Note that only one frequency of each type of excitation in the hierarchy (optical, microwave and radiofrequency) is required, simplifying the experimental arrangement.

\section{Conclusion}
We have described a procedure for graph state quantum computing that is tailored to fully exploit the physics of optically active multi-level systems. We assign to each physical system two logical qubits: the {\em broker} and the {\em client}. Groups of brokers negotiate new graph state fragments via an optical mechanism that can be prone to failure. Completed fragments are mapped from broker to clients via a simple state transition and measurement, thus the clients remain entirely insulated from failures during the brokering. This has the consequence that arbitrary graph topologies can be grown efficiently even when the brokerage failure rate is high. 

There are many possible physical systems possessing a multi-level spectrum that can be exploited by this approach. We describe an implementation in terms of \nv~centres in diamond, where brokers and clients are very naturally embodied as electron and nuclear spins. Demonstrator experiments appear to be feasible immediately. 



\section{Acknowledgments}
\begin{acknowledgments}
The authors thank Jason Smith and Jens Eisert for helpful discussions and comments on the manuscript. SCB is supported by the Royal Society. DEB is supported by Merton College, Oxford and by the QIPIRC. JF acknowledges funding from the Embark Iniative. JJLM acknowledges support from EPSRC.
\end{acknowledgments}



\begin{thebibliography}{99}

\bibitem{blossom}
Searching the quant-ph online archive for papers containing the terms ``cluster state'' or ``graph state''
in their abstract reveals that the numbers have roughly doubled each year for the last three years.

\bibitem{BandR}
	H.J. Briegel and R. Raussendorf, 
	Phys. Rev. Lett. \textbf{86} 910 (2001).
	
\bibitem{RandB}
	R. Raussendorf and H.J. Briegel, 
	Phys. Rev. Lett. \textbf{86} 5188 (2001).

\bibitem{Long}
        R.\ Raussendorf, D.E.\ Browne, and H.J.\ Briegel,
        Phys. Rev. A {\bf 68}, 022312 (2003).

\bibitem{Graphs}
        M.\ Hein, J.\ Eisert, and H.J.\ Briegel,
        Phys. Rev. A {\bf 69}, 062311 (2004).

\bibitem{nielsenLOQC} 
	M.A. Nielsen, 
	Phys. Rev. Lett. {\bf 93}, 040503 (2004).

\bibitem{Jaksch} 
	S.R. Clark, C. Moura Alves, D. Jaksch, 
	New J. Phys. {\bf 7}, 124 (2005).

\bibitem{browneLOQC} 
	D.E. Browne and T. Rudolph, 
	Phys. Rev. Lett. {\bf 95}, 010501 (2005).

\bibitem{Kay} 
	A. Kay, J.K. Pachos, C.S. Adams, 
	quant-ph/0501166.

\bibitem{BK}
        S.D.\ Barrett and P.\ Kok, 
        Phys.\ Rev.\ A {\bf  71}, 060310(R) (2005).

\bibitem{BenjComment} 
        S.C.\ Benjamin,
        quant-ph/0504111.

\bibitem{Almut}
        Y.-L.\ Lim, A.\ Beige, and L.C.\ Kwek,
Phys.\ Rev.\ Lett.\ {\bf 95}, 030505 (2005).

\bibitem{BES}
        S.C. Benjamin, J. Eisert and T.M. Stace,
        New J. Phys {\bf 7}  194 (2005).

\bibitem{LimLong}
        Y.L.Lim {\em at al.}, 
        quant-ph/0508218.
        
\bibitem{BellLambda}
	http://www.bell-labs.com/news/1999/november/10/1.html


\bibitem{grover}
	L.K. Grover, quant-ph/960724.

\bibitem{jens1}
        J.I. Cirac, A.K. Ekert, S.F. Huelga, and C. Macchiavello, 
        Phys. Rev. A {\bf 59}, 4249 (1999).

\bibitem{jens2}
        J. Eisert, K. Jacobs, P. Papadopoulos and M.B. Plenio,
        Phys. Rev. A {\bf 62}, 052317 (2000).

\bibitem{sandu}
	D. Collins, N. Linden, and S. Popescu,
	Phys. Rev. A {\bf 64}, 032302 (2001).

\bibitem{ionTrapCoupling}
        L.-M. Duan, B.B. Blimov, D.L. Moehring and C. Monroe,
        Quant. Inf. Comput. {\bf 4}, 165 (2004).

\bibitem{jelezko04}
        F. Jelezko and T. Gaebel and I. Popa and A. Gruber and J. Wrachtrup,
        Phys. Rev. Lett. {\bf 92}, 076401 (2004).


\bibitem{Cabrillo} C. Cabrillo {\em et al}, Phys. Rev. A {\bf 59}, 1025 (1999).

\bibitem{Bose KPV 99} S. Bose {\em et al},
Phys. Rev. Lett. {\bf 83}, 5158 (1999).

\bibitem{brownehuelga} D.E. Browne, S.F. Huelga and M.B. Plenio, Phys. Rev. Lett. \textbf{91} 067901 (2003). 

\bibitem{polarization}  
X-L Feng {\em et al}, Phys. Rev. Lett. {\bf 90} 217902 (2003); L.-M. Duan and H. J. Kimble, Phys. Rev. Lett. \textbf{90},253601 (2003); C. Simon and W.T.M. Irvine Phys. Rev. Lett. \textbf{91} 110405 (2003). 

\bibitem{entgates}  I.E. Protsenko {\em et al}, Phys. Rev. A, {\bf 66}, 062306 (2002); J. Cho and H.-W. Lee, quant-ph/0509005.


\bibitem{hemansonprb} X.-F. He, Neil B. Manson and Peter T.H. Fisk, Phys. Rev. B \textbf{47} 8809 (1993). 

\bibitem{hemmeroptlett} P.R. Hemmer \textit{et al}, Opt. Lett. {26} 361 (2001).


\bibitem{verstraete} F. Verstraete and J. I. Cirac, Phys. Rev. A 70, 060302(R) (2004).

\bibitem{losstolerant} M. Varnava, D.E. Browne and T. Rudolph, quant-ph/0507036.

\bibitem{faulttolerant} C.M. Dawson, H.L. Haselgrove and M.A. Nielsen, quant-ph/0509060; P. Aliferis and D.W. Leung quant-ph/0503130; M.A. Nielsen and C.M Dawson Phys. Rev. A 71, 042323 (2005).




\bibitem{ladd05}
        T. D. Ladd and D. Maryenko and Y. Yamamoto and E. Abe and K. M. Itoh,
        Phys. Rev. B {\bf 71}, 014401 (2005).

\bibitem{wrachtrup01}
        J. Wrachtrup and S. Y. Kilin and A. P. Nizovtsev,
        Optics and Spectroscopy, {\bf 91}, 459 (2001).

\bibitem{cox94}
        A. Cox and M. E. Newton and J. M. Baker,
        J. Phys. Cond. Mat. {\bf 6}, 551 (1994).

\bibitem{he92}
        X-F. He and P. T. H. Fisk and N. B. Manson,
        J. Appl. Phys. {\bf 72}, 211 (1992).


\end{thebibliography}
\end{document}